\documentclass[]{aastex631}

\usepackage{subfigure}

\shorttitle{kinematic analysis of Patchick 99}
\shortauthors{Butler et al.}

\graphicspath{{./}{figures/}}

\usepackage{rotating}
\begin{document}

\title{RR Lyrae Stars Belonging to the Candidate Globular Cluster Patchick 99}

\author[0000-0002-1533-6004]{Evan Butler}
\affiliation{Department of Physics, University of Washington, Physics-Astronomy Bldg, Room C121, Box 351560, Seattle, WA, 98195-1560}
\affiliation{Saint Martin's University, 5000 Abbey Way SE, Lacey, WA, 98503, USA}

\author[0000-0002-2808-1370]{Andrea Kunder}
\affiliation{Saint Martin's University, 5000 Abbey Way SE, Lacey, WA, 98503, USA}

\author[0000-0001-5497-5805]{Zdenek Prudil}
\affiliation{European Southern Observatory, Karl-Schwarzschild-Strasse 2, 85748 Garching bei M\"{u}nchen, Germany}

\author[0000-0001-6914-7797]{Kevin R. Covey}
\affiliation{Department of Physics \& Astronomy, Western Washington University, MS-9164, 516 High St., Bellingham, WA, 98225}

\author{Macy Ball}
\affiliation{Saint Martin's University, 5000 Abbey Way SE, Lacey, WA, 98503, USA}

\author[0009-0009-4825-429X]{Carlos Campos}
\affiliation{Saint Martin's University, 5000 Abbey Way SE, Lacey, WA, 98503, USA}

\author{Kaylen Gollnick}
\affiliation{Department of Physics \& Astronomy, Western Washington University, MS-9164, 516 High St., Bellingham, WA, 98225}

\author{Julio Olivares Carvajal}
\affiliation{Instituto de Astrof\'{i}sica, Pontificia Universidad Cat\'{o}lica de Chile, Av. Vicu\~{n}a Mackenna 4860, 782-0436 Macul, Santiago, Chile}
\affiliation{Millennium Institute of Astrophysics, Av. Vicu\~{n}a Mackenna 4860, 82-0436 Macul, Santiago, Chile}

\author{Joanne Hughes}
\affiliation{Physics Department Seattle University, 901 12th Ave., Seattle, WA 98122, USA}

\author[0000-0002-3723-6362]{Kathryn Devine}
\affiliation{The College of Idaho, 2112 Cleveland Blvd Caldwell, ID, 83605, USA}

\author[0000-0002-8878-3315]{Christian I. Johnson}
\affiliation{Space Telescope Science Institute, 3700 San Martin Drive, Baltimore, MD 21218, USA}

\author[0000-0003-4341-6172]{A.~Katherina~Vivas}
\affiliation{Cerro Tololo Inter-American Observatory/NSF’s NOIRLab, Casilla 603, La Serena, Chile}

\author[0000-0003-0427-8387]{Michael R. Rich}
\affiliation{Department of Physics and Astronomy, UCLA, 430 Portola Plaza, Box 951547, Los Angeles, CA 90095-1547, USA}

\author[0000-0002-8717-127X]{Meridith Joyce}
\affiliation{Konkoly Observatory, HUN-REN Research Centre for Astronomy and Earth Sciences, Konkoly-Thege Mikl\'os \'ut 15-17, H-1121, Budapest, Hungary}
\affiliation{CSFK, MTA Centre of Excellence, Budapest, Lend\"ulet Near-Field Cosmology Research Group, 1121, Budapest, Konkoly-Thege Mikl\'os \'ut 15-17, H-1121, Budapest, Hungary}

\author[0000-0001-8889-0762]{Iulia T. Simion}
\affiliation{Shanghai Key Lab for Astrophysics, Shanghai Normal University, 100 Guilin Road, Shanghai, 200234}

\author{Tommaso Marchetti}
\affiliation{European Southern Observatory, Karl-Schwarzschild-Strasse 2, 85748 Garching bei M\"{u}nchen, Germany}

\author[0000-0002-9859-4956]{Andreas J. Koch-Hansen}
\affiliation{Zentrum f\"ur Astronomie der Universit\"{a}t Heidelberg, Astronomisches Rechen-Institut, M\"{o}nchhofstr. 12-14, 69120 Heidelberg, Germany}

\author[0000-0002-2577-8885]{William I. Clarkson}
\affiliation{Department of Natural Sciences, University of Michigan-Dearborn, 4901 Evergreen Rd. Dearborn, MI 48128, USA}

\author{Rebekah Kuss}
\affiliation{Saint Martin's University, 5000 Abbey Way SE, Lacey, WA, 98503, USA}
\affiliation{Department of Mathematics, Oregon State University, 1500 SW Jefferson Way, Corvallis, OR 97331}

\begin{abstract}

Patchick~99 is a candidate globular cluster located in the direction of the Galactic bulge, with a proper motion almost identical to the field and extreme field star contamination.  
A recent analysis suggests it is a low-luminosity globular cluster with a population of RR Lyrae stars.  
We present new spectra of stars in and around Patchick~99, targeting specifically the 3 RR Lyrae stars associated with the cluster as well as the other RR Lyrae stars in the field. 
A sample of 53 giant stars selected from proper motions and a position on CMD are also observed.  
The three RR Lyrae stars associated with the cluster have similar radial velocities and distances, and two of the targeted giants also have radial velocities in this velocity regime and $\rm [Fe/H]$ metallicities that are slightly more metal-poor than the field.  
Therefore, if Patchick~99 is a bonafide globular cluster, it would have a radial velocity of $-$92$\pm$10~km~s$^{-1}$, a distance of 6.7$\pm$0.4~kpc (as determined from the RR Lyrae stars), and an orbit that confines it to the inner bulge.  
\end{abstract}


\section{Introduction} \label{sec:intro}

The globular clusters surrounding our Milky Way are nearly as old as the Universe itself.   
These roughly spherical collections of hundreds of thousands of tightly packed and gravitationally bound stars form in the early Universe in high-density peaks \citep{diemand05, boley09}.  
Their high density allows them to endure tidal disintegration, and they, therefore, witness most of the formation and evolution processes of galaxies, which they can be used to study \citep{brodie06}. 

In contrast to globular clusters in the halo, clusters in the inner Galaxy are more likely to in-spiral toward the galactic center due to the stronger tidal field in the inner Galaxy and dynamical friction.  
It is believed that clusters inside the central 2~kpc of the Milky Way have lost $\sim$80\% of their initial population \citep{baumgardt18}.  
Therefore, low-luminosity/low-mass GCs should exist in the inner Galaxy and should likely be more ubiquitous in the inner Galaxy than the halo.
However, to date, there is a factor of $\sim$5 times more low-luminosity GCs known in the halo of the Milky Way \citep{baumgardt21}.  

One survey searching for potential low-luminosity candidate globular clusters in the inner Galaxy is the VISTA Variables in the Via-Lactea survey (VVV) survey, a large ($\sim$1~billion star) photometric, infrared time domain survey of the disk and inner Galaxy \citep{minniti10}.  
Between 2011 and 2018, more than 100 new GC candidates in the bulge have been proposed, 
most of them (but not all) have low luminosities.  
After {\it Gaia} proper motions were released in 2018, it could be seen if ``cluster" stars were moving together in proper motion space \citep[e.g.,][]{gran19}. However, stellar spectroscopy is still necessary to identify individual star cluster members from the field and to validate candidate star clusters as being separate from the field \citep[some examples of this include][]{fernandeztrincado21, gran22}.

This study focuses on the low-luminosity candidate GC Patchick~99 at ($l$,$b$)=(2.4884\textdegree, $-$6.1452\textdegree).  
Identified by Dana Patchick on 20 November 2004 using 2MASS imagery, the Deepskyhunters group \citep[e.g.,][]{kronberger06} recognized it first as `DSH J1815.7-2948'(D. Patchick, priv. communication). Since then, it has appeared in internal catalogs as Patchick~99.        
\citet{bica19} list it as a globular cluster in their compilation of star clusters, associations, and candidates in the Milky Way. 
It was the subject of a study by \citet{garro21}, who postulate Patchick~99 is a bonafide globular cluster based on {\it Gaia} proper motions, color-magnitude diagrams (CMDs) from VVV photometry, and by identifying a population of RR Lyrae stars (RRLs) that were thought to be associated with Patchick~99.  
The extreme similarity in both the proper motion and the CMD between the field and the potential cluster stars makes identification of this cluster difficult, and no radial velocity measurements of this potential cluster exist to date.

The first spectroscopic observations are presented here in an attempt to probe the nature of Patchick~99, particularly by targeting its RR Lyrae star population.

\section{Observations and Data Reductions} \label{sec:data}

\begin{figure}
\centering
\mbox{\subfigure{\includegraphics[width=10cm]{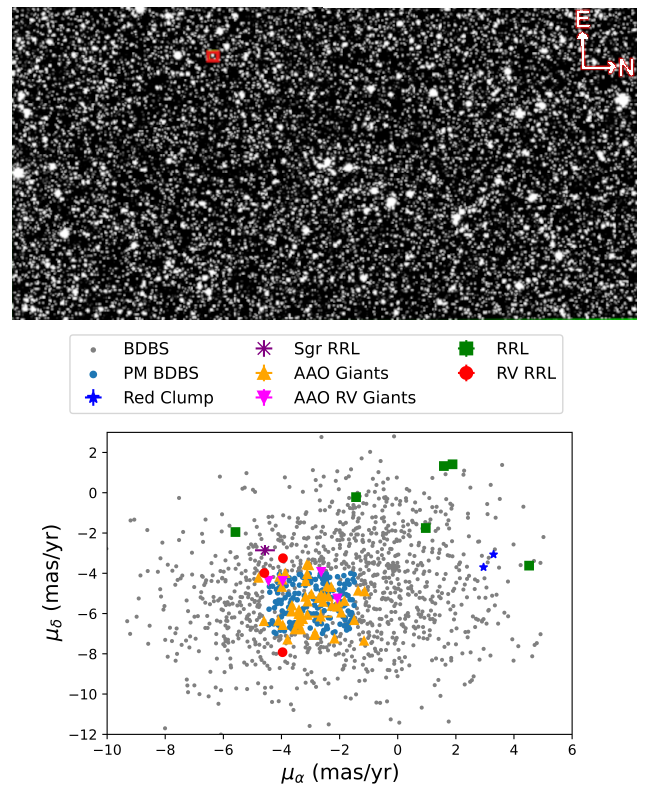}}
\subfigure{\includegraphics[width=7cm]{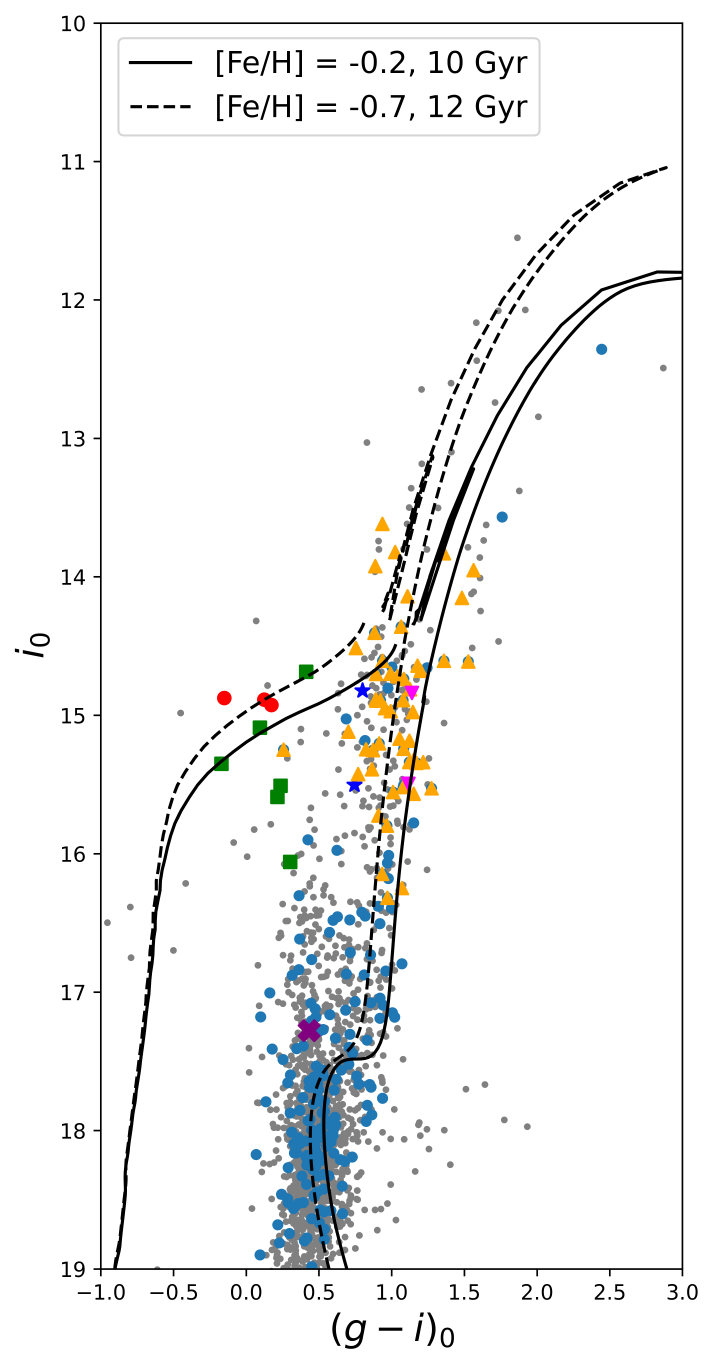}}}
\caption{
{\it Left, top:} A BDBS $z$-band image of Patchick~99 is shown with dimensions of 9' x 4.5';  OGLE-BLG-RRLYR-35459, one of the three RRLs associated with this cluster, is highlighted.
{\it Left, bottom:} This proper motion distribution shows spectroscopically targeted stars presented here (bolded) as compared to BDBS stars within 2.4 arcminutes of Patchick~99 (grey). 
The most probable members of Patchick~99 are the RRLs designated with circles (red) as well as the giants designated with upside-down triangles (magenta).
{\it Right:} The BDBS color-magnitude diagram of stars within 2.4' of Patchick~99 (grey) as compared to the spectroscopically targeted stars presented here (bolded).
}
\label{pms}
\end{figure}

\subsection{Observations and Target Selection} 
The observations come from the AAOmega multifiber spectrograph on the Anglo-Australian Telescope (PROP-ID: O/2022A/3002) with the red 1700D grating centered at 8600~\AA (so that the calcium II triplet 
lines were observed) giving a resolution of R$\sim$11,000.
Exposure times ranged from 4x30 min to 2x30 min, adjusting for weather and seeing conditions.
The data reduction steps were carried out using the AAOmega 2dfdr software: Bias subtraction, quartz-flatfielding, cosmic ray cleaning, sky subtraction using 35 designated sky fibers, and wavelength calibration via arc-lamp exposures. 
The typical final spectrum had a wavelength range from 8350--8800 \AA, with slight variations depending on the spectrum's exact position on the CCD.

In total, 53 giants, 41 red clump stars, and 271 RRLs within a 2~degree field of view of Patchick~99 were targeted. 
The selected giants had {\it Gaia} DR3 proper motions within ($\mu_\alpha$, $\mu_\delta$)=($-$2.98~mas~yr$^{-1}\pm$2.5~mas~yr$^{-1}$, $-$5.49~mas~yr$^{-1}\pm$2.5~mas~yr$^{-1}$) \citep[the mean proper motion of Patchick~99 as presented in][]{garro21, gaiacolab23} as well as a RUWE \textless1.2.
This is a larger proper motion range than the proper motion uncertainty presented in \citet{garro21} to encompass all possible members.  
All 12 RRLs within 10 arcminutes of the cluster were observed, regardless of proper motion.  
Red clump stars from \citet{johnson22} with photometric $\rm [Fe/H]$ metallicities more metal-poor than $-$0.25~dex were included as the lowest priority targets in the configuration files. 
The giants were only observed once---each AAOmega plate configuration retained the same RRLs and red clump stars but cycled through different giant stars.  

The focus here is the stars around the central 10~arcminutes of Patchick~99, for which the center of $\alpha$=273.94583\textdegree and $\delta$=$-$29.81278\textdegree presented in \citet{garro21} is used. 
Figure~\ref{pms} (lower left panel) shows the proper motion distribution of the observed stars 
as compared to the underlying field observed in the Blanco DECam Bulge Survey \citep[BDBS,][]{rich20, johnson20}.
BDBS is a large optical survey of the southern Galactic bulge in the $ugrizY$ filters spanning ~200 square degrees from $-$11\textdegree \textless l \textless $+$11\textdegree and $-$13\textdegree \textless b \textless $-$2\textdegree \citep[BDBS,][]{rich20, johnson20}.
To avoid the foreground disk, all targeted stars had {\it Gaia} parallaxes $\leq$0.4~mas \citep{marchetti22}.

The targeted giants all had $griz$ photometry from BDBS that indicated they followed the color-magnitude diagram (CMD) of the potential Patchick~99 CMD.  In Figure~\ref{pms} (right panel), the dereddened $i_0$ versus $(g-i)_0$ CMD is shown, where the reddening procedure from \citet{simion17} was used to account for extinction along the line-of-sight.
Two possible MIST \citep[MESA Isochrones and Stellar Tracks,][]{choi16} isochrones that adopt the cluster's distance derived in this work (6.7$\pm$0.4~kpc)---one with $\rm [Fe/H]$=$-$0.2 and one with $\rm [Fe/H]$=$-$0.7---are shown.  
The horizontal branch (HB) isochrones come from PGPUC\footnote{http://www2.astro.puc.cl/pgpuc/iso.php} using DECam filters, and standard GC values for a metal-rich bulge cluster, $e.g.$, helium ($Y$=0.25), $\rm [\alpha/Fe] = 0.2$, progenitor mass = 0.8$M_\odot$ \citep{valcarce12, gran21}.

The BDBS CMD shows the isochrone using the cluster metallicity and age as put forward by \citet{garro21} ($\rm [Fe/H] = -$0.2, age=10~Gyr).  We also show the isochrone using the cluster metallicity determined here and an age more indicative of bulge GCs with RRLs ($\rm [Fe/H] = -$0.7, age=12~Gyr).  The optical CMD presented here is not used to constrain any cluster parameters but to verify consistency between the BDBS photometry and the cluster parameters derived here using RRLs.

\subsection{Velocities}
IRAF's {\tt xcsao} routine was used to calculate radial velocities by cross-correlating science spectra against five calibration spectra.
These calibration templates were chosen from the Apache Point Observatory Galaxy Evolution Experiment \citep[APOGEE DR17,][]{abdurro22} DR17 database and all observed during our 5-night run at the Anglo-Australian Telescope. 
Specifically, the radial velocity templates used were from APOGEE~2M18134674-2926056 (RV=27.88$\pm$0.03~km~s$^{-1}$), APOGEE~2M17514997-2906055  (RV=$-$187.33$\pm$0.02~km~s$^{-1}$), APOGEE~2M17521244-2919510 (RV=65.13$\pm$0.05~km~s$^{-1}$), APOGEE~2M17525012-3146525 (RV=$-$221.609$\pm$0.02~km~s$^{-1}$), and APOGEE~2M17505621-3240401 (RV=$-$229.373$\pm$0.10~km~s$^{-1}$).  

\begin{figure}
\centering
\mbox{\subfigure{\includegraphics[width=4.5cm]{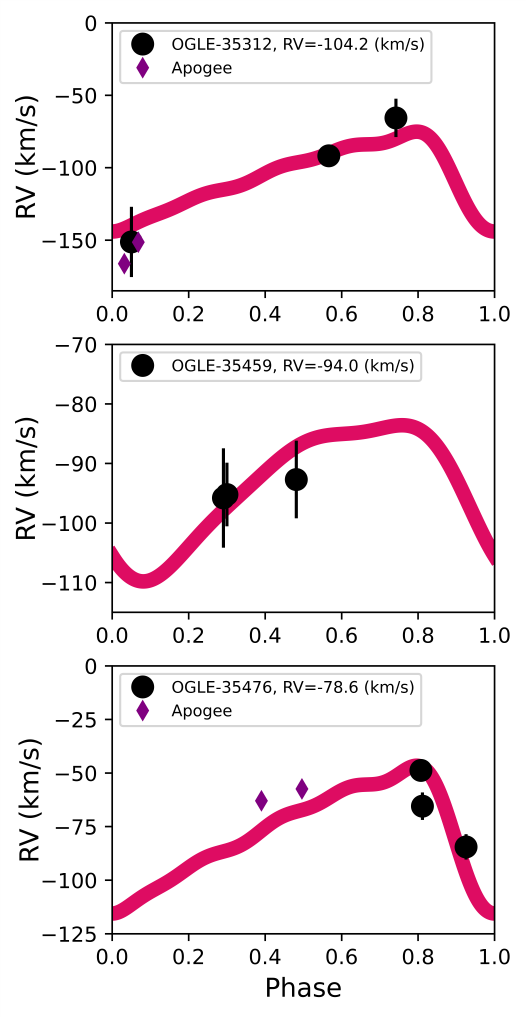}}\subfigure{\includegraphics[width=8.9cm]{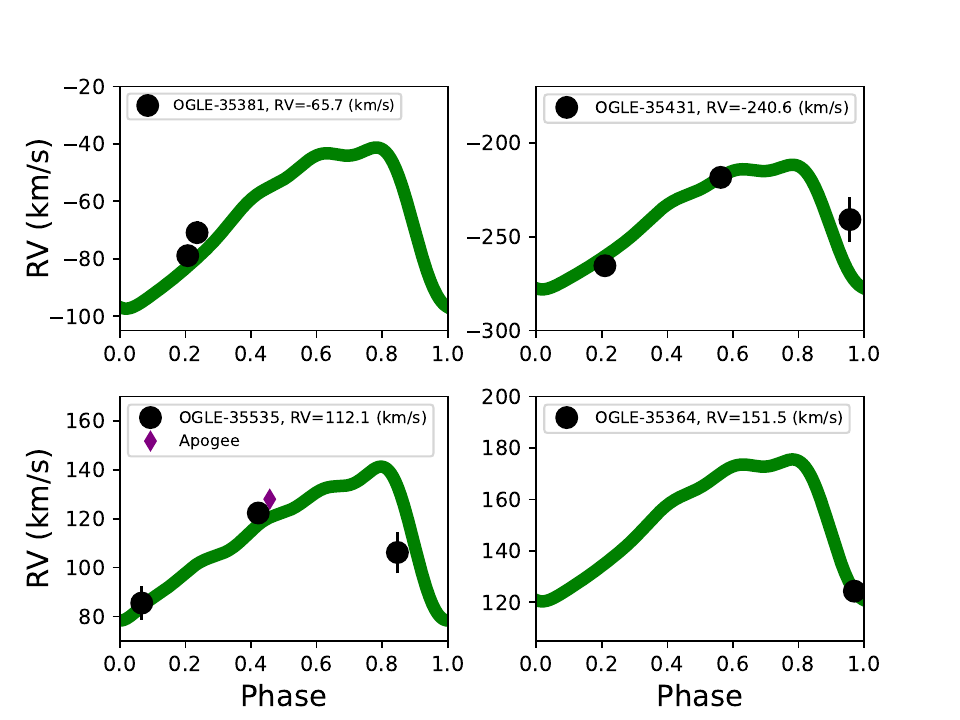}}\subfigure{\includegraphics[width=4.5cm]{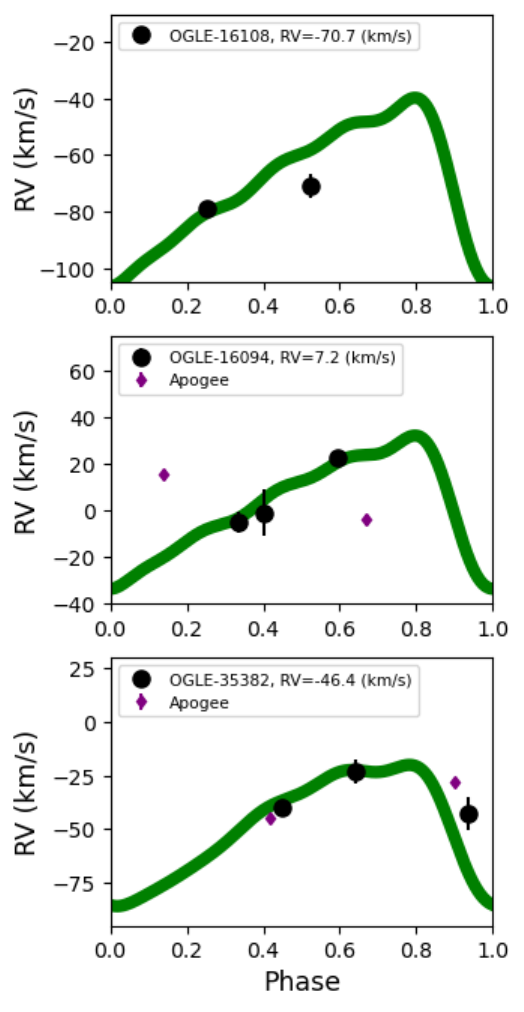}}}
\caption{The line-of-sight velocity curves of ten RRLs within 10~arcmin of Patchick~99, where our observations are designated by solid circles (black).
The \citet{prudil24a} templates shown are used to determine the systemic velocities for the RRLs.
The three leftmost panels (red) indicate the RRLs with {\it Gaia} proper motions consistent with Patchick~99. 
The four center and rightmost RRL velocity curves do not have proper motions consistent with Patchick~99.  
Five stars have APOGEE observations, which are designated by diamonds (purple).
}
\label{lightcurves}
\end{figure}

Twelve RRLs---ten RRab stars and two RRc stars---within 10~arcminutes of Patchick~99's center were observed three times during the observing run.  The velocity curves for 10 of these RRLs are shown in Figure~\ref{lightcurves}.  
The spectrum for the RRab star OGLE-BLG-RRLYR-35355 had a signal-to-noise per pixel of $\leq$2 and so was excluded from kinematic analysis.  
The RRc star OGLE-BLG-RRLYR-35447 was also excluded from analysis, as the extracted spectrum was a superposition of the two different stars landing in the AAOmega fiber, and we were unable to separate the two spectra to isolate that of the RRL.

The RRL velocity curves show the corresponding template from \citet{prudil24a}, derived specifically for RRLs in the bulge.  Each observation time was converted into pulsation phase, $\phi$, using the OGLE pulsation ephemerides and pulsation periods so that maximum light falls at $\phi$=0 for each RRL \citep[OGLE,][]{udalski15}.
These templates were used to calculate center-of-mass radial velocities for the stars, using the Fourier fits from the template spectra to fit the template model to the observations \citep{prudil24a}. 
The uncertainty adopted for each star is 10 km~s$^{-1}$, which includes the uncertainties in the 
individual radial velocity measurements combined with the uncertainties in the template fitting.

Five of the ten stars were observed as part of the APOGEE DR17. 
To overplot these onto the radial velocity curves, the time of observation specified by the column {\tt JD} 
in the APOGEE {\tt allVisit-dr17-synspec\_rev1.fits}{\footnote{https://www.sdss4.org/dr17/irspec/spectro\_data/} file was used.}
There is good agreement between the APOGEE observations and our observations, except for the star OGLE-BLG-RRLYR-16094, where the APOGEE observations appear to be offset in phase by $\sim$0.35.  This could be due to a varying period of this star, although this star is not flagged as a period-changing star or Blazhko variable 
in the OGLE catalog \citep{prudil17}.

We searched the {\it Gaia} DR3 database \citep{gaiacolab23} for additional stars with proper motions, radial velocities, photometric metallicities, and photometric distances consistent with Patchick 99. Apart from the observed giants, we were not successful in finding any star in {\it Gaia} that we believed was consistent with Patchick 99 membership.

\section{Patchick~99 Discussion}

\subsection{Velocities}
Six RRLs are postulated to be Patchick~99 cluster members by \citet{garro21} based on their period-luminosity
relation.  However, only three of those have proper motions consistent with Patchick~99.  
The mean heliocentric radial velocity of these three stars is $-$92.3$\pm$6~km~s$^{-1}$. 
To take into account the random errors of $\pm$10~km~s$^{-1}$ in the RRL systemic velocities, the intrinsic velocity dispersion, $\sigma_0^2$ = $\sigma_{vel}^2$ - $\sigma_{err}^2$, is found. The obtained internal dispersion of $\sigma_0$=3~km~s$^{-1}$ is typical for globular clusters, especially in the lower luminosity range.  

The rotation curve of the Galactic bulge is well known \citep[e.g.,][]{kunder12}, and at the location of Patchick~99, the typical galactocentric velocity is $\sim$30~km~s$^{-1}$.  
In contrast, the mean galactocentric velocity of the three Patchick~99 RRLs is $\sim -$75~km~s$^{-1}$, indicating their velocities are distinct from the underlying bulge field. 

Four giants are found to be within the radial velocity range of the RRLs ($i.e.$, between $-$107 and 
$-$75~km~s$^{-1}$).  
These are listed in Table~\ref{table}.  
We show in \S\ref{sec:metallicity} that two of those four giants, the two with radial velocities $\sim -$100~km~s$^{-1}$, have spectroscopic metallicities that are consistent with belonging to Patchick~99.  The two giants with radial velocities of $\sim -$80~km~s$^{-1}$ are not as metal-poor as the RRL and therefore more likely to belong to the bulge field.
The two red clump stars have neither a consistent proper motion nor radial velocity and are shown in Figure~\ref{pms} and Figure~\ref{RV_FeH}. 

Table~\ref{table} lists the radial velocities of all the RR~Lyrae stars, as well as the giants and red clump stars.  
A portion of the table is shown for clarity; all 53 observed giants are included in the electronic version of this paper.

\subsection{Metallicity} \label{sec:metallicity}
\begin{figure}
\centering
\mbox{\subfigure{\includegraphics[width=18cm]{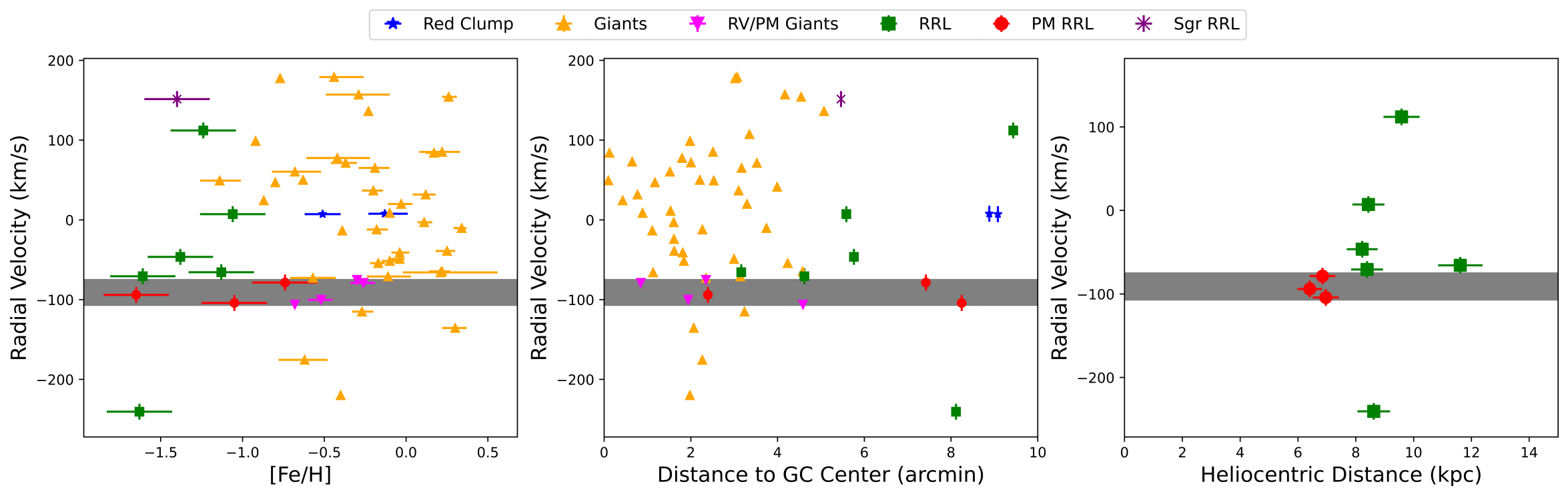}}}
\caption{
{\it Left:} The photometric and spectroscopic metallicities and heliocentric radial velocities of stars within 10~arcminutes of Patchick~99's center. 
The radial velocities of all RR Lyraes have an uncertainty of $\sim$10~km~s$^{-1}$.
{\it Center:} The heliocentric radial velocities of stars within 10~arcminutes of Patchick~99.
{\it Right:} The heliocentric distances for the RR Lyrae stars within 10~arcminutes of Patchick~99's center.
{\it All:} RR Lyrae stars with radial velocities and proper motions consistent with Patchick~99 are marked with red circles.
Giants with radial velocities and proper motions consistent with Patchick~99 are marked with magenta upside-down triangles.
}
\label{RV_FeH}
\end{figure}

The photometric metallicities for the observed RRab stars are calculated using the relations from \citet{dekany21} and are listed in Table~\ref{table} and shown in Figure~\ref{RV_FeH}. The formal error in these metallicities is $\pm$0.2 \citep{dekany21}, although photometric metallicities derived from the pulsation properties are better suited to describe the average metallicity of a population of RRLs rather than the metal abundance of an individual RRL.
The two RRab candidate Patchick~99 RRLs have photometric $\rm [Fe/H]$ metallicities that are the two most metal-rich stars of the RRL sample, with $\rm [Fe/H]$ = $-$1.04 and $\rm [Fe/H]$ = $-$1.35 on the \citet{for11, chadid17, sneden17, crestani21} metallicity scale, abbreviated as CFCS.  This translates to $\rm [Fe/H]$ = $-$0.74 and $\rm [Fe/H]$ = $-$1.05 on the APOGEE ASPCAP $\rm [Fe/H]$ metallicity scale \citep{kunder24} which we will use throughout this analysis.  
%
The photometric metallicity for the RRc candidate Patchick~99 RRL is $\rm [Fe/H]$ = $-$1.65. 
Photometric metallicities for RRc stars are not as robust as for the RRab stars as the spectroscopic calibrating set for the RRc stars is 1/3 the size of the spectroscopic calibrating set for the RRab stars and has especially few stars at the higher metallicity range.  

Due to the similarity in the CMD between Patchick~99 and the bulge field, it was postulated that this cluster is on the metal-rich end.  The photometric metallicities of the RRab stars do suggest it has a metallicity more metal-rich than the typical bulge RRL metallicity, which peaks at $[Fe/H] = -$1.2~dex \citep{dekany21}, but we find it is more metal-poor than the $\rm [Fe/H] = -$0.2~dex estimate put forward by \citet{garro21} from CMD fitting.  

The SP\_ACE code \citep{boeche21} was used to determine spectroscopic metallicities for both the giants and red clump stars in our sample.  
\citet{kunder24} use AAOmega spectra taken with the same wavelength range and resolution used here to show that SP\_ACE can reproduce $\rm [Fe/H]$ metallicities with an accuracy of 0.2~dex over a metallicity range of $\rm [Fe/H]\sim -$0.9 to $\sim$ +0.2~dex. 

Two of the four giants with radial velocities similar to the RRL in Patchick~99 have $\rm [Fe/H]$ metallicities within one sigma of the Patchick~99 RRLs. These two stars are more metal-poor than the majority of the field stars and have an average $\rm [Fe/H]$ metallicity of $-$0.60.  
These are the most likely candidate Patchick~99 giants.  

\subsection{Distance and Orbit}\label{sec:distance}

Distances to the RRLs were determined using empirical period–absolute magnitude–metallicity (PMZ) relations from \citet{prudil24b}, which are calibrated from local RRLs with {\it Gaia} parallaxes.  
These are intended to be used in conjunction with the {\it Gaia} catalog and the bulge photometric surveys of OGLE and VVV and are therefore particularly well suited for the RRLs presented here.  
Figure~\ref{RV_FeH} (right panel) shows the distances of 9 of the 10 RRLs observed; OGLE-BLG-RRLYR-35364 is not shown, as it is at a distance of $\sim$26~kpc.  This distance and radial velocity are consistent with OGLE-BLG-RRLYR-35364 belonging to the Sagittarius Dwarf Galaxy \citep{kunderchab09}. 

The three candidate RRLs of Patchick~99 cluster in distance space, with distances within one sigma of each other.
Their mean heliocentric distance is $d_\odot$=6.7$\pm$0.4~kpc, which is similar to the distance of 6.6$\pm$0.6~kpc reported by \citet{garro21}.
The other RR Lyrae stars within 10' from Patchick~99 have distances offset from the candidate RRL cluster stars.

\begin{figure}
\centering
\mbox{\subfigure{\includegraphics[width=13.2cm]{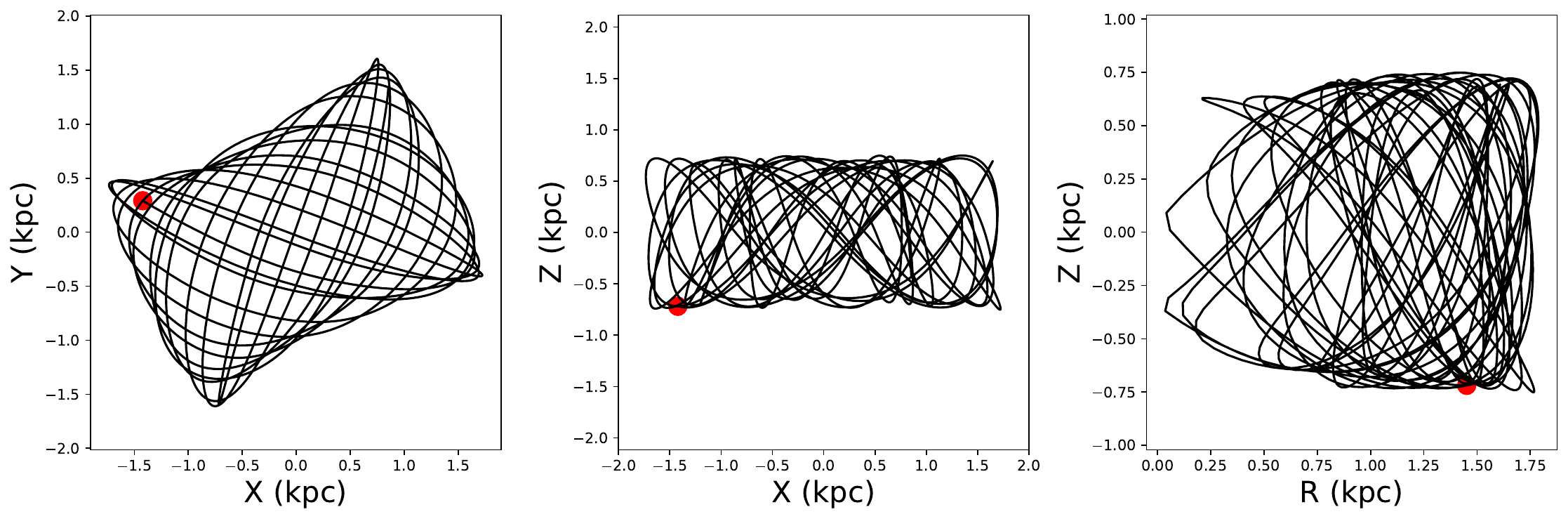}}}
\caption{
The integrated orbit of Patchick~99 using as initial conditions the distance and radial velocity presented here along with the position and proper motion values from \citet{garro21}. The projection of the orbit on the Galactic plane (x - y), the (x - z) plane, and the meridional plane (R - z) are all shown and suggest this candidate cluster is confined to the inner Galaxy. The cluster's present-day location is highlighted as a red dot.}
\label{orbit}
\end{figure}

To simulate the orbital trajectory of Patchick 99 over the past Gyr, we used the Gala python package \citep{price-whelan2017, price-whelan2023}.  For this calculation, we adopt a Galactic potential with three components: a flattened 6x10$^{10}$ M$_{\odot}$ \citet{MiyamotoNagai1975} stellar disk with a 3.5~kpc scale length and 0.280~kpc scale height; a tri-axial \citet{LongMurali1992} bar with M=10$^{10}$ M$_{\odot}$ and appropriate shape parameters (a=4~kpc, b=0.8~kpc, c=0.25~kpc, and $\alpha$=25 degrees); and a 6x10$^{11}$ M$_{\odot}$ \citet{NFW1996} dark matter halo with a 20~kpc scale radius. 
The cluster's present-day orbital parameters were defined with the values given earlier in this paper: namely, the cluster’s position (RA = 273.94583$\circ$, Dec = $-$29.81278$\circ$), proper motion ($\mu_\alpha$=$-$2.98$\pm$2.5~mas~yr$^{-1}$, $\mu_\delta$=$-$5.49$\pm$2.5~mas~yr$^{-1}$), radial velocity ($-$92$\pm$10~km~s$^{-1}$) and distance (6.7$\pm$0.4~kpc).  
By defining a timestep of $-$0.5 Myr and then integrating the orbit over 2000 steps, the last billion years of Patchick 99's orbit were calculated and shown in Figure~\ref{orbit}.  
During this period, the cluster has been in a flattened orbit that does not extend beyond 1.75~kpc from the Galactic center or 0.75~kpc from the Galactic midplane. 
We have experimented with changing the distance, radial velocity, and proper motion within their uncertainties and also experimented with changing the Galactic potential. 
Although this does affect the specific shape of the orbit, the cluster always remains confined to the innermost 2~kpc of the Galactic center.

\section{Conclusions and Final Remarks}\label{sec:conclusion}

Patchick~99 is a candidate low-luminosity GC located in the inner Galaxy. 
We present the first spectroscopic observations for stars within 10' of the center of Patchick~99 to determine the radial velocity of stars belonging to this structure.  Our focus is on the RR Lyrae stars because giants are ubiquitous in the bulge field and heavily overlap with both the color-magnitude diagram and proper motion distribution of Patchick~99 \citep{garro21}.  The use of RRLs as probes of globular clusters is advantageous as overdensities of horizontal branch stars can separate globular clusters from the crowded bulge field \citep[e.g.][]{rich20}. 

The three RRLs with proper motions consistent with the cluster have similar radial velocities of $\langle RV \rangle$=$-$92.3$\pm$10~km~s$^{-1}$ with a standard deviation of $\sigma$=3~km~s$^{-1}$. They further clump in distance space, with a mean heliocentric distance of $d_\odot$=6.7$\pm$0.4~kpc.
Therefore, the RRL population of Patchick~99 suggests it is a bonafide low-luminosity GC on the near side of the bulge.

A sample size of three is not ideal for determining the mean properties of a cluster; having a larger sample of stars would improve the mean velocity and dispersion for this potential cluster.  
The low luminosity of Patchick~99 means only a handful of giant and HB stars belonging to the cluster will exist, and identifying cluster members will be challenging.  
For example, \citet{simpson17} found only 11 cool giants and HB stars in the low-luminosity cluster ESO~452-SC11 (cluster mass of 6.8$\pm$3.4~x~10$^3 M_{\odot}$).  
In this study, very few giants targeted have properties indicating they could be probable cluster members.  In particular, two of the 53 giants observed have velocities and $\rm[Fe/H]$ metallicities consistent with cluster membership.  Probing the fainter main-sequence stars may be needed for future studies.
The metallicities of these two giants combined with the photometric metallicities of the RRLs indicate a mean metallicity of $-$0.75$\pm$0.3~dex.  
The metallicity and distance found here is consistent with isochrones fit to the BDBS CMD.

Given the low luminosity of Patchick~99, it is worth considering how likely it would be for this cluster to harbor a population of three RRLs.  Synthetic HB procedures may be usefully adopted to predict relevant features of the frequency of RRLs as a function of cluster mass, but this is beyond the scope of our paper.
Observationally, it is worth noting that ultra-faint dwarf galaxies with similar brightness as Patchick~99 have 1--12 RRLs \citep{martinezvazquez19, vivas20}. 

The small mass of Patchick~99 combined with its position just a few kiloparsecs of the Galactic Center means it must have lost most of its initial mass and should be close to final dissolution \citep{baumgardt03}.  The spread in radial velocity and proper motion of the RRLs may indicate that Patchick~99 is a fragment of a larger tidally disrupted structure rather than a globular cluster. 
Additional observations focused on the region immediately surrounding Patchick~99 are necessary to better constrain the nature and characteristics of Patchick~99.

\clearpage
\begin{sidewaystable}
\begin{scriptsize}
\centering
\begin{tabular}{|p{3.2cm}|p{2.1cm}|p{2.1cm}|p{1.7cm}|p{1.4cm}|p{2.1cm}|p{2.1cm}|p{2.2cm}|p{1.8cm}|p{1cm}|}
    \hline
    Source-ID & R.A.~(\textdegree) & Dec~(\textdegree) & $d_\odot$~(kpc) & r\textsubscript{P99}~(') & $\mu_\alpha$~(mas~yr$^{-1}$) & $\mu_\delta$~(mas~yr$^{-1}$) & HRV~(km~s$^{-1}$) & $\rm [Fe/H]$ & Type \\
    \hline \hline
    OGLE-BLG-RRLYR-35476$^{\star}$ & 274.0053$\pm$0.0347 & $-$29.9212$\pm$0.0314 & 6.85$\pm$0.44 & 7.420 & $-$4.59$\pm$0.04 & $-$3.99$\pm$0.03 & $-$78.6$\pm$10 & $-$0.74$\pm$0.20 & RRab \\
    \hline
    OGLE-BLG-RRLYR-35312$^{\star}$ & 273.8098$\pm$0.0369 & $-$29.8325$\pm$0.0318 & 6.96$\pm$0.45 & 8.247 & $-$3.95$\pm$0.05 & $-$3.25$\pm$0.03 & $-$104.2$\pm$10 & $-$1.05$\pm$0.20 & RRab \\
    \hline
    OGLE-BLG-RRLYR-35459$^{\star}$ & 273.9749$\pm$0.0603 & $-$29.8401$\pm$0.0563 & 6.41$\pm$0.43 & 2.392 & $-$3.96$\pm$0.08 & $-$7.92$\pm$0.06 & $-$94.0$\pm$10 & $-$1.65$\pm$0.20 & RRc\\
    \hline
    \hline
    4049645773449540864 & 273.9069$\pm$0.0446 & $-$29.8169$\pm$0.0398 & -- & 2.349 & $-$3.96$\pm$0.05 & $-$4.38$\pm$0.04 & $-$75.26$\pm$1.84 & $-$0.3$\pm$0.20 & Giant\\
    \hline
    4049646013961775232 & 273.9553$\pm$0.0360 & $-$29.8025$\pm$0.0323 & -- & 0.846 & $-$4.45$\pm$0.04 & $-$4.37$\pm$0.03 & $-$79.04$\pm$1.35 & $-$0.26$\pm$0.28 & Giant\\
    \hline
    4049646254479960832$^{\star}$ & 273.9205$\pm$0.0404 & $-$29.7930$\pm$0.0363 & -- & 1.940 & $-$2.63$\pm$0.05 & $-$3.92$\pm$0.04 & $-$100.20$\pm$1.16 & $-$0.52$\pm$0.27 & Giant\\
    \hline
    4049649419971997952$^{\star}$ & 273.9137$\pm$0.1023 & $-$29.7437$\pm$0.0867 & -- & 4.585 & $-$2.08$\pm$0.13 & $-$5.25$\pm$0.10 & $-$106.20$\pm$1.83 & $-$0.68$\pm$0.20 & Giant\\
    \hline
    \multicolumn{10}{|c|}{Observed stars that do not have velocities consistent with Patchick~99} \\
    \hline
    OGLE-BLG-RRLYR-16094 & 273.9599$\pm$0.0369 & $-$29.7208$\pm$0.0324 & 8.43$\pm$0.55 & 5.586 & 4.52$\pm$0.05 & $-$3.62$\pm$0.03 & 7.2$\pm$10 & $-$1.06$\pm$0.20 & RRab\\
    \hline
    OGLE-BLG-RRLYR-35382 & 273.8947$\pm$0.0372 & $-$29.8940$\pm$0.0334 & 8.22$\pm$0.53 & 5.760 & 1.89$\pm$0.04 & 1.41$\pm$0.03 & $-$46.4$\pm$10 & $-$1.38$\pm$0.20 & RRab\\
    \hline
    OGLE-BLG-RRLYR-16108 & 273.9885$\pm$0.0471 & $-$29.7487$\pm$0.0409 & 8.39$\pm$0.55 & 4.618 & 1.59$\pm$0.06 & 1.33$\pm$0.04 & $-$70.7$\pm$10 & $-$1.61$\pm$0.20 & RRab\\
    \hline
    OGLE-BLG-RRLYR-35381 & 273.8937$\pm$0.0511 & $-$29.8050$\pm$0.0458 & 11.62$\pm$0.77 & 3.163 & $-$1.44$\pm$0.06 & $-$0.22$\pm$0.05 & $-$65.7$\pm$10 & $-$1.13$\pm$0.20 & RRab\\
    \hline
    OGLE-BLG-RRLYR-35535 & 274.0875$\pm$0.0722 & $-$29.8810$\pm$0.0604 & 9.58$\pm$0.63 & 9.433 & $-$5.58$\pm$0.09 & $-$1.95$\pm$0.07 & 112.1$\pm$10 & $-$1.24$\pm$0.20 & RRab\\
    \hline
    OGLE-BLG-RRLYR-35431 & 273.9381$\pm$0.0410 & $-$29.9478$\pm$0.0382 & 8.62$\pm$0.56 & 8.115 & 0.97$\pm$0.05 & $-$1.75$\pm$0.04 & $-$240.6$\pm$10 & $-$1.63$\pm$0.20 & RRab\\
    \hline
    OGLE-BLG-RRLYR-35364 & 273.8741$\pm$0.2738 & $-$29.8689$\pm$0.2809 & 26.26$\pm$2.58 & 5.461 & $-$4.57$\pm$0.34 & $-$2.86$\pm$0.26 & 151.5$\pm$10 & $-$1.40$\pm$0.20 & Sgr RRab\\
     \hline
    4049646701162462208 & 273.8062$\pm$0.0520 & $-$29.8618$\pm$0.0474 & -- & 8.881 & 2.95$\pm$0.06 & $-$3.70$\pm$0.05 & 7.7$\pm$10 & $-$0.13$\pm$0.12 & RC\\
    \hline
    4049739678702776064 & 274.0793$\pm$0.0378 & $-$29.7415$\pm$0.0342 & -- & 9.080 & 3.30$\pm$0.04 & $-$3.06$\pm$0.03 & 7.2$\pm$10 & $-$0.51$\pm$0.11 & RC\\
    \hline
\end{tabular}
\caption{\label{table}This table is published in its entirety in the electronic version of this paper. A portion is shown here for clarity. The asterisk (*) symbol indicates stars with velocities and metallicities consistent with Patchick~99.}
\end{scriptsize}
\end{sidewaystable}

\begin{acknowledgments}
AMK acknowledges support from grant 
AST-2009836 from the National Science Foundation. AMK, KRC, JH, KD acknowledge the M.J. Murdock Charitable Trust's support through its RAISE (Research Across Institutions for Scientific Empowerment) program.  
This work was made possible through the Preparing for
Astrophysics with LSST Program, supported by the Heising-Simons Foundation and
managed by Las Cumbres Observatory.  
M.J. gratefully acknowledges the funding of MATISSE: \textit{Measuring Ages Through Isochrones, Seismology, and Stellar Evolution}, awarded through the European 
Commission's Widening Fellowship. This project has received funding from the European Union's Horizon 2020 research and innovation programme.

Based in part on data obtained at Siding Spring Observatory [via PROP-ID: O/2022A/3002]. 
We acknowledge the traditional owners of the land on which the AAT stands, the Gamilaraay people, and pay our respects to elders past and present.

This project used data obtained with the Dark Energy Camera (DECam), which was constructed by the Dark Energy Survey (DES) collaboration. Funding for the DES Projects has been provided by the US Department of Energy, the US National Science Foundation, the Ministry of Science and Education of Spain, the Science and Technology Facilities Council of the United Kingdom, the Higher Education Funding Council for England, the National Center for Supercomputing Applications at the University of Illinois at Urbana-Champaign, the Kavli Institute for Cosmological Physics at the University of Chicago, Center for Cosmology and Astro-Particle Physics at the Ohio State University, the Mitchell Institute for Fundamental Physics and Astronomy at Texas A\&M University, Financiadora de Estudos e Projetos, Fundação Carlos Chagas Filho de Amparo à Pesquisa do Estado do Rio de Janeiro, Conselho Nacional de Desenvolvimento Científico e Tecnológico and the Ministério da Ciência, Tecnologia e Inovação, the Deutsche Forschungsgemeinschaft and the Collaborating Institutions in the Dark Energy Survey.

The Collaborating Institutions are Argonne National Laboratory, the University of California at Santa Cruz, the University of Cambridge, Centro de Investigaciones En\'{e}rgeticas, Medioambientales y Tecnol\'{o}gicas–Madrid, the University of Chicago, University College London, the DES-Brazil Consortium, the University of Edinburgh, the Eidgen\"{o}ssische Technische Hochschule (ETH) Z\"{u}rich, Fermi National Accelerator Laboratory, the University of Illinois at Urbana-Champaign, the Institut de Ci\'{e}ncies de l’Espai (IEEC/CSIC), the Institut de Física d’Altes Energies, Lawrence Berkeley National Laboratory, the Ludwig-Maximilians Universität München and the associated Excellence Cluster Universe, the University of Michigan, NSF’s NOIRLab, the University of Nottingham, the Ohio State University, the OzDES Membership Consortium, the University of Pennsylvania, the University of Portsmouth, SLAC National Accelerator Laboratory, Stanford University, the University of Sussex, and Texas A\&M University.

Based on observations at Cerro Tololo Inter-American Observatory, NSF’s NOIRLab (NOIRLab Prop. ID 2013A-0529; 2014A-0480; PI: M. Rich), which is managed by the Association of Universities for Research in Astronomy (AURA) under a cooperative agreement with the National Science Foundation.

This work used data from the Sloan Digital Sky Survey (SDSS). 
Funding for the Sloan Digital Sky Survey IV has been provided by the Alfred P. Sloan Foundation, the U.S. Department of Energy Office of Science, and the Participating Institutions. 

SDSS-IV acknowledges the support and resources from the Center for High-Performance Computing at the University of Utah. The SDSS website is www.sdss4.org.

SDSS-IV is managed by the Astrophysical Research Consortium for the Participating Institutions of the SDSS Collaboration including the Brazilian Participation Group, the Carnegie Institution for Science, Carnegie Mellon University, Center for Astrophysics | Harvard \& Smithsonian, the Chilean Participation Group, the French Participation Group, Instituto de Astrof\'isica de Canarias, The Johns Hopkins University, Kavli Institute for the Physics and Mathematics of the Universe (IPMU) / University of Tokyo, the Korean Participation Group, Lawrence Berkeley National Laboratory, Leibniz Institut f\"ur Astrophysik Potsdam (AIP),  Max-Planck-Institut f\"ur Astronomie (MPIA Heidelberg), Max-Planck-Institut f\"ur Astrophysik (MPA Garching), Max-Planck-Institut f\"ur Extraterrestrische Physik (MPE), National Astronomical Observatories of China, New Mexico State University, New York University, University of Notre Dame, Observat\'ario Nacional / MCTI, The Ohio State University, Pennsylvania State University, Shanghai Astronomical Observatory, United Kingdom Participation Group, Universidad Nacional Aut\'onoma de M\'exico, University of Arizona, University of Colorado Boulder, University of Oxford, University of Portsmouth, University of Utah, University of Virginia, University of Washington, University of Wisconsin, Vanderbilt University, and Yale University.

This work has made use of data from the European Space Agency (ESA) mission {\it Gaia} (\url{https://www.cosmos.esa.int/gaia}), processed by the {\it Gaia} Data Processing and Analysis Consortium (DPAC, \url{https://www.cosmos.esa.int/web/gaia/dpac/consortium}). Funding for the DPAC has been provided by national institutions, in particular, the institutions participating in the {\it Gaia} Multilateral Agreement.

\end{acknowledgments}

{}

\end{document}